\documentclass[
]{ceurart}

\usepackage[most]{tcolorbox}
\usepackage{xcolor}
\usepackage{geometry} 
\usepackage{float}
\definecolor{promptback}{rgb}{0.96, 0.96, 0.96}  
\definecolor{promptframe}{rgb}{0.80, 0.80, 0.80} 

\lstdefinestyle{promptstyle}{
  basicstyle=\itshape,    
  breakatwhitespace=false,
  breaklines=true,        
  captionpos=b,           
  extendedchars=true,     
  showspaces=false,       
  showstringspaces=false, 
  showtabs=false,         
  escapeinside={\%*}{*)}, 
  literate={- \*}{--}2
}

\newtcolorbox{researchpromptbox}[1][]{
  colback=promptback,     
  colframe=promptframe,    
  arc=0mm,                
  boxrule=0.5pt,          
  breakable,              
  enhanced,               
  title={\textbf{Prompt template used in the Semantic LLM-based method}},
  toptitle=1.5mm,         
  bottomtitle=1.5mm,
  coltitle=black,         
  colbacktitle=promptback,
  sharp corners=downhill, 
  listing only,           
  listing engine=listings,
  listing options={style=promptstyle}, 
  #1                      
}

\usepackage{listings}
\usepackage[all]{nowidow}
\usepackage{tabularx}
\begin{document}

\copyrightyear{2026}
\copyrightclause{Copyright for this paper by its authors.
  Use permitted under Creative Commons License Attribution 4.0
  International (CC BY 4.0).}


\title{When AI Meets Science: Research Diversity, Interdisciplinarity, Visibility, and Retractions across Disciplines in a Global Surge}



\author[1,2]{Andrés F. Castro-Torres}[
orcid=0000-0003-1032-3869, 
email= andres.castro@bsc.es]

\author[1]{Joan Giner-Miguelez}[%
orcid=0000-0003-2335-6977,
email=joan.giner@bsc.es
]

\author[1]{Mercè Crosas}[%
orcid=0000-0003-1304-1939,
email=merce.crosas@bsc.es
]

\address[1]{Computational Social Science and Humanities, Barcelona Supercomputing Center, BSC-CNS, Barcelona, Spain}
\address[2]{Max Planck Institute for Demographic Research, Rostock, Germany}

\conference{Preprint article}
\begin{abstract}
The extent to which Artificial Intelligence (AI) technologies can trigger generalized paradigm shifts in science is unclear. Although these technologies have revolutionized data collection and analysis in specific fields, their overall impact depends on the scope and ways of adoption. We analyze over 227 million scholarly works from the OpenAlex collection (1960-2024) spanning four scientific domains and 46 fields. To distinguish the use of AI as research method (AI adoption) from mentioning AI-related terms (AI engagement), we developed a two-step AI-assisted semantic classification pipeline, validated through human coding of 911 abstracts and a robustness check on 348,000 full-text articles (PLOS One). We document differences in the timing and extent of AI adoption across domains, with generalized exponential growth after 2015. The transformative nature of this growth, however, is less apparent. AI-supported research is confined to a few topics with strong ties to Computer Science and conventional statistical frameworks, suggesting limited epistemological transformation. It is also associated with an unwarranted citation premium and substantially higher retraction rates than non-AI-supported. Geographically, while wealthy countries lead in AI publications per capita, global South countries in a belt from Indonesia to Algeria lead in AI adoption relative to their national output, signaling a distinctive resource concentration pattern. The transformative capacity of AI in science thus remains untapped, and its rapid adoption underlines challenges in research openness, transparency, reproducibility, and ethics. We discuss how best research practices could boost the benefits of AI adoption and highlight areas that warrant closer scrutiny.

\end{abstract}

\begin{keywords}
  Artificial Intelligence \sep
  Scientific Discovery \sep
  Global Science \sep
  Research Integrity \sep
  Bibliometrics 
\end{keywords}

\maketitle

\section{Introduction}

Artificial Intelligence (AI) technologies are influencing multiple sectors of society, including scientific research \cite{mitchell_2020}, driving a rapid increase in AI developments, adoption, and debate within academia. The existing literature, however, presents mixed arguments on the potential of AI for scientific innovation and discovery, with perspectives that range from enthusiasm and cautious optimism to skepticism, criticism, dissent, refutation, and outright rejection \cite{guest_2025_uncritical, channing_2025_aiscientific_preprint, sourati_2023_accelerating, messeri_2024_illusions, bender_2021_stochastic, swanson_2025_virtuallab, channing_2026_aiscientific}. The 2024 Nobel Prize in Chemistry, awarded to Hassabis, Jumper, and Baker for breakthroughs in protein structure prediction and design, epitomized the optimistic pole of this debate and amplified expectations about AI's widespread adoption in research \cite{wang_2023_scientific}. More skeptical voices, however, have highlighted the potential pitfalls of AI adoption, including power imbalances vis-à-vis big tech corporations, the dehumanizing consequences of some AI technologies (primarily so-called generative AI), the ethical aspects of data collection and use for model training, and the environmental implications of the infrastructure that supports these technologies \cite{bender_2025_aicon, crawford_2021_atlas, denton_2021_genealogy}. 

\begin{figure}[b]
  \centering
  \label{fig:engagement}
  \includegraphics[width=0.9\linewidth]{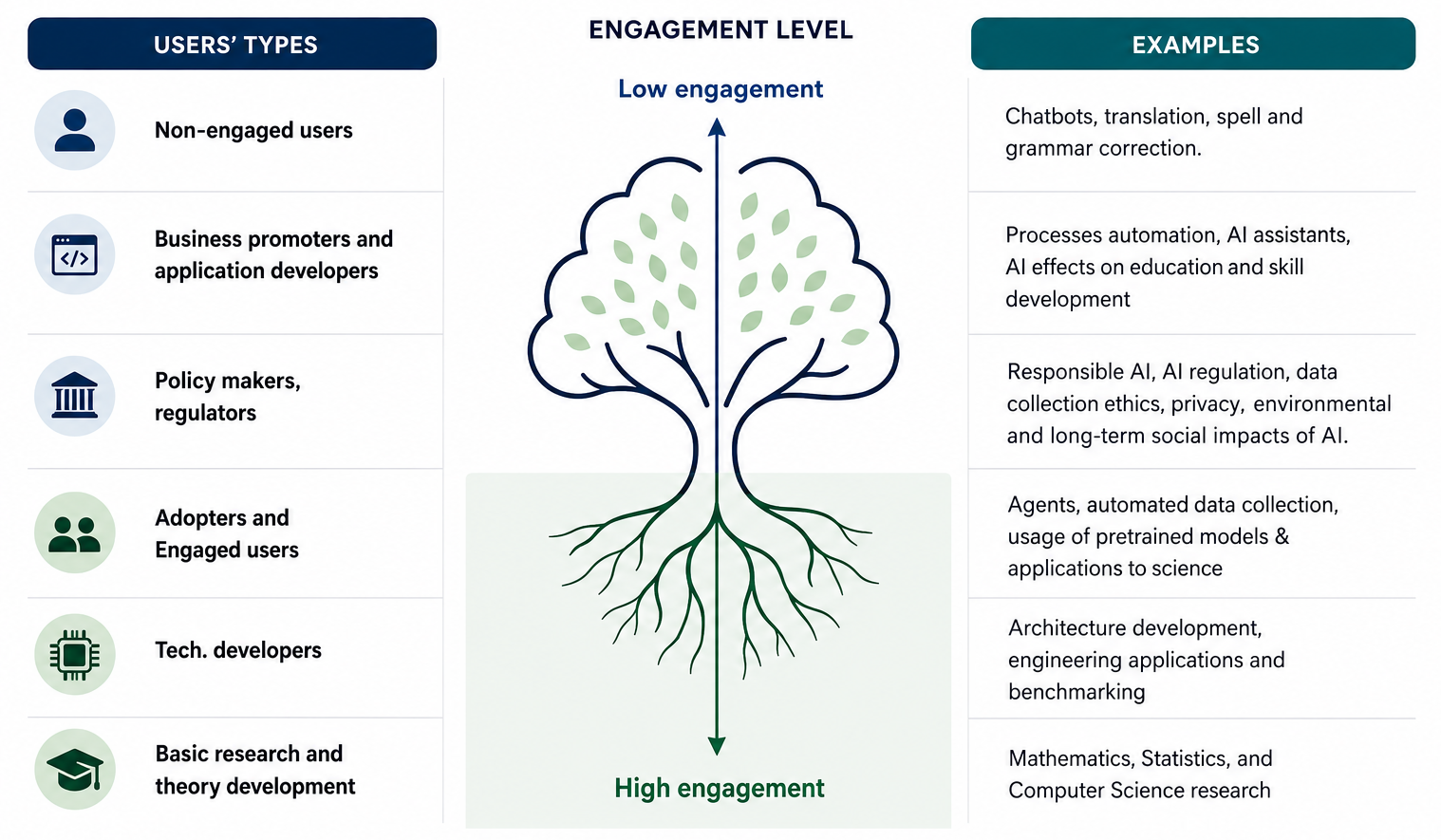}
  \caption{Artificial Intelligence users’ typology along engagement levels and use examples. The tree represents the collection of AI technologies, including its roots on basic research and technology development which nurture AI adoption in science represented by the “Adopters and Engaged users” category. The tree’s stem represents policy makers, government, and other regulatory institutions which we deem crucial for developing responsible AI-based products and services offered by multiple companies (tree’s branches). In contrast to researchers and technology developers (tree’s roots), end users’ engagement with AI, defined as the level of understanding of how the supporting AI systems, models and methods work, is minimum. The two-sided color of the tree represents contrasting views across all user types, from rejection to critiques to heightened optimism, that are very apparent in existing literature, social media, and public discussions.}
  \label{fig:feature}
\end{figure}

The lack of a widely accepted definition of what AI is, and the fact that the term has been used to group a multiplicity of things from automatic translation tools, classification and clustering methods to expert systems, and more recently, Large Language Models (LLM), further complicates the conversion  \cite{guest_2025_critical}. As depicted metaphorically in Figure \ref{fig:engagement}, AI technologies are diverse from roots to stems, branches, and fruits implying different user types and diverging levels of engagement from developers to end users. In this proposed conceptualization of AI technologies, AI adoption in science corresponds to the underneath part of the tree mainly driven by AI developments and applications for data collection, processing, and analysis, all of which imply a deep understanding of the theoretical, technical and social implications of AI applications. Importantly, critical perspectives, represented by the three's split appearance, spread along the engagement distribution from critical adopters and critical users including those who reject specific uses of these technologies, speaking of AI’s susceptibility to create contrasting views at all levels.

The fundamental role played by AI in the 2024 Nobel Prize in Chemistry, awarded to Hassabis, Jumper, and Baker for breakthroughs in protein structure prediction and design, epitomized the potential of AI for science and it increased the already heightened optimism about its widespread adoption in research \cite{wang_2023_scientific}. More skeptical voices, however, have also highlighted the potential pitfalls and shortcomings of AI adoption including the influence and power imbalances vis a vis big tech corporations, the dehumanizing consequences of some AI technologies (primarily the so-called generative AI), as well as the ethical aspects of data collection and usage for models’ training and the environmental implications of the infrastructure that supports AI technologies \cite{bender_2025_aicon, crawford_2021_atlas, denton_2021_genealogy}. 

The lack of a widely accepted definition of what AI is, and the fact that the term has been used to group a multiplicity of things from automatic translation tools, classification and clustering methods to expert systems, and more recently, Large Language Models (LLM), further complicates the conversion \cite{guest_2025_critical}. As depicted metaphorically in Figure \ref{fig:engagement}, AI technologies are diverse from roots to stems, branches, and fruits implying different user types and diverging levels of engagement from developers to end users. In this proposed conceptualization of AI technologies, AI adoption in science corresponds to the underneath part of the tree mainly driven by AI developments and applications for data collection, processing, and analysis, all of which imply a deep understanding of the theoretical, technical and social implications of AI applications. Importantly, critical perspectives, represented by the two-side colors, spread along the engagement distribution from critical adopters and critical users including those who reject specific uses of these technologies, speaking of AI’s susceptibility to create contrasting views at all levels.

Despite these contrasting views and diverse engagement levels, scholars with both enthusiastic and critical perspectives on AI would agree that its most durable impacts in science are likely to come from adopters and engaged users, as the Nobel prize case has shown. Data collection and analysis are stages where the development and adoption of AI technologies are more likely to augment scholars' capacity, potentially opening new research areas by allowing them to ask new questions or provide new answers to existing questions. However, AI adoption for these purposes can also lead to narrower research if scholars start choosing research topics based overwhelmingly on its opportunity to use AI, expecting to gain visibility due to the salience of the technology rather than based on pertinence, interest, and topics’ relevance criteria \cite{hao_2024_aitools}. 

In addition, AI adoption for data collection and analysis has two potential side effects in science. First, it could contribute to widening existing knowledge production inequalities across countries and world regions. This is because the financial, technical, and material resources required to develop and adopt these technologies are concentrated in wealthy nations \cite{crawford_2021_atlas}. Second, as AI technologies continue to figure prominently in society, scholars will engage in AI discussions, meaning that researchers will study the causes and consequences of AI developments and adoption in science. 

This study documents and discusses temporal and geographical trends of AI adoption in science across four scientific domains and 46 fields from 1960 to 2024. This historical perspective and field disaggregation is possible thanks to the massive nature of our data, the Open Alex collection. This collection gives us access to the bibliometric information of over 227 million academic works. We compared AI-based scientific works (i.e., scientific works that use AI as part of their data collection and analysis) vs. non-AI-based scientific works in terms of topical concentration, reliance on existing statistical methods, citations to Computer Science works, received citations, and retraction rates (i.e., number of retracted papers per 1,000 publications within a given field). These analyses show that, compared to non-AI-based research, AI-based research is: (i) more concentrated in a few topics; (ii) more likely to rely on existing statistical analysis frameworks (i.e., generalized linear model frameworks); (iii) more prone to cite works from Computer Sciences, but not works from other fields; (iv) more prone to receive citations; and (v) more likely to be retracted. Some scientific fields within the Physical Science and Engineering, including Computer Sciences, are exempted from this latter pattern as retraction rates between AI-based and non-AI-based works do not differ. Finally, we show country-level heterogeneities in the rate of AI adoption in science, as the number of AI-based articles per population, and the prevalence of AI in science, as the percentage of AI-based articles per 100 publications.

We use these findings to highlight how existing best practices and protocols in science, including data management, open science protocols, and research sharing schemas (FAIR), should be used to harvest the benefits of increased AI adoption. At the same time, we highlight the challenges that AI technologies imply for greater research openness, transparency, and ethics. Our findings also uncover scientific fields and geographical areas where AI adoption (or its lack thereof) require a closer look due to exceptional trends (e.g., high levels and rapid growth of AI adoption), potentially problematic patterns (e.g., greater retraction rates and unwarranted greater citations), their potential relation with the gender and geographical dimension of knowledge production (e.g.., where and who produces AI-based research), and geographic- and field-specific trends that may be of interest to governments, and research funding institutions.

\section{What We Know about AI in Science}

According to our proposed conceptualization (Figure \ref{fig:engagement}), scholars use AI systems, models, and methods (AI technologies hereafter) for different purposes and with different levels of engagement. For example, on the one hand, using a chatbot for spellchecking, literature search, or coding are examples of an instrumental use of AI technologies with low engagement and likely negligible epistemological implications. On the other hand, using Neural Networks for image classification in medical research, developing clustering algorithms for engineering applications, using LLMs for text analysis, or discussing the implication of any of these methods for science, are examples of highly engaged AI use with the potential to enhance scientific discovery and innovations, i.e., with potentially relevant epistemological implications. 

Moreover, among highly engaged AI uses, we distinguish three forms:: (i) the promotion of AI discussions, i.e., scholars that take part in conversations about the use, development, and adoption of AI technologies in science without necessarily using them; (ii) the engagement in AI adoption, i.e., scholars that use AI technologies to collect and analyze data as part of their research strategies; and (iii) contributing AI developments, i.e., scholars that create new or improve existing AI technologies regardless of their potential applications in research. This latter category includes hardware, software, and fundamental research developers.

These three forms of relation to AI are not mutually exclusive and straightforward to disentangle. Scholars may relate to AI in more than one form simultaneously, e.g., as critics and developers \cite{mitra_2025_emancipatory}, as developers and adopters \cite{do_2024_augmented}, or as adopters and critics \cite{ollion_2024_dangers}. For example, AI adoption and development are hard to disentangle in problem-oriented research because the development of AI applications is in itself a research question, which mixes features of development and adoption in research articles. This is likely the case of research in the Physical and Engineering Sciences. Likewise, works that adopt or develop AI could likely include AI discussions, making these two categories not necessarily mutually exclusive.

Previous studies have investigated scholarly uses of AI by searching AI-related terms in articles’ titles and abstracts; they term this AI engagement. Once AI-engaged articles are identified, authors have analyzed articles’ distribution across fields and journals, and the semantic similarity among AI-engaged and non-AI-engaged articles \cite{duede_2024_oilwater, hajkowicz_2023_artificial}. According to these studies, AI engagement is growing and spreading rapidly, meaning that the number of publications using AI-related words has increased exponentially, with a 1293\% growth between 1985 and 2022. AI-engaged publications have reached a fairly large share of journals beyond those specialized in AI topics in terms of citations and as publication outlets. This rapid growth has been accompanied by decreased semantic heterogeneity, whereby works using AI-related terminology are less heterogeneous than works that do not include AI-related terms as measured by the cosine of works’ embeddings \cite{duede_2024_oilwater}. Using similar word-search approaches, other studies have linked AI use with authors’ sociodemographic characteristics, showing the existence of gender and race/ethnicity differential in AI use/engagement \cite{gao_2024_quantifying}.

Due to the word-search approach, it remains unclear whether this growth and integration of AI into the scientific literature is due to increased AI discussion, adoption, and development. In all these cases, AI-related terms are likely to appear in works’ titles and abstracts, and cross-domain citations should be expected in articles that discuss, adopt, or develop AI technologies, albeit for different reasons in each case. 

Another limitation of these studies is that AI-use prevalence and growth are measured with respect to the entire body of academic works. This approach mixes articles that may not be susceptible to methods of empirical analysis. There are several academic works that do not necessarily tackle empirical questions, but rather theoretical or conceptual issues, which makes them less susceptible to being AI adopters in the first place. 

This bias has been partially solved by studies that focus on a few scientific fields where presumably AI could be generally applied, as the share of works that could rely on AI is high \cite{gao_2024_quantifying, hao_2024_aitools}. According to these studies, the potential effects of AI in research are transformative, and caution is needed to properly harvest the benefits of AI while preventing potential harms, particularly with regard to research ethics, replicability, and resource reuse \cite{ginermiguelez_2025_readiness, leist_2022_mapping, hosseini_2025_openscience}. Yet, selecting fields does not provide a general picture of AI adoption and neglects the real extent of domain- and field-specific heterogeneities in AI adoption. 

Relatedly, the literature with a critical stand against certain aspects of AI technologies more broadly in society is growing rapidly. This literature include works on the ethics of data collection for AI and AI usage \cite{ricaurte_2022_ethics, tacheva_2023_aiempire}, their environmental impact \cite{valdivia_2025_supply}, the potential consequences for social and technological inequalities \cite{wang_2023_scientific}, their impact on higher education systems \cite{guest_2025_uncritical}, and the perils of unwarranted anthropomorphization of AI tools (\cite{placani_2024_anthropomorphism}). This rapid growth further undermines the usefulness of word-search-based approaches as they lump together works engaged with AI adoption and development with works engaged in AI discussions, from supportive to critical ones. This more critical literature on AI lacks, albeit from a different angle, the exploration of heterogeneous patterns across fields, as well as the nuances across different forms of relation and use of AI, particularly in terms of the adoption of AI technologies as research methods.

Due to the relatively reduced scope of the data used in previous studies, the geography of AI adoption and the bibliometric characteristics of AI-based research at a global scale remains poorly documented despite the widespread recognition of the key role of geopolitics and world-level economic relations to these technologies. Studying the bibliometric characteristics of AI-based research could help understanding the implications of AI adoption and adapt existing best research practices to the specificities of working with AI technologies. Likewise, documenting geographical patterns will serve to measure the scope of AI adoption, identify leaders and laggers, and consider the role of AI in existing knowledge production inequalities.

\section{Data and Methods}

\subsection{Architecture of the data pipeline}

\begin{figure}
  \centering
  \label{fig:method}
  \includegraphics[width=1\linewidth]{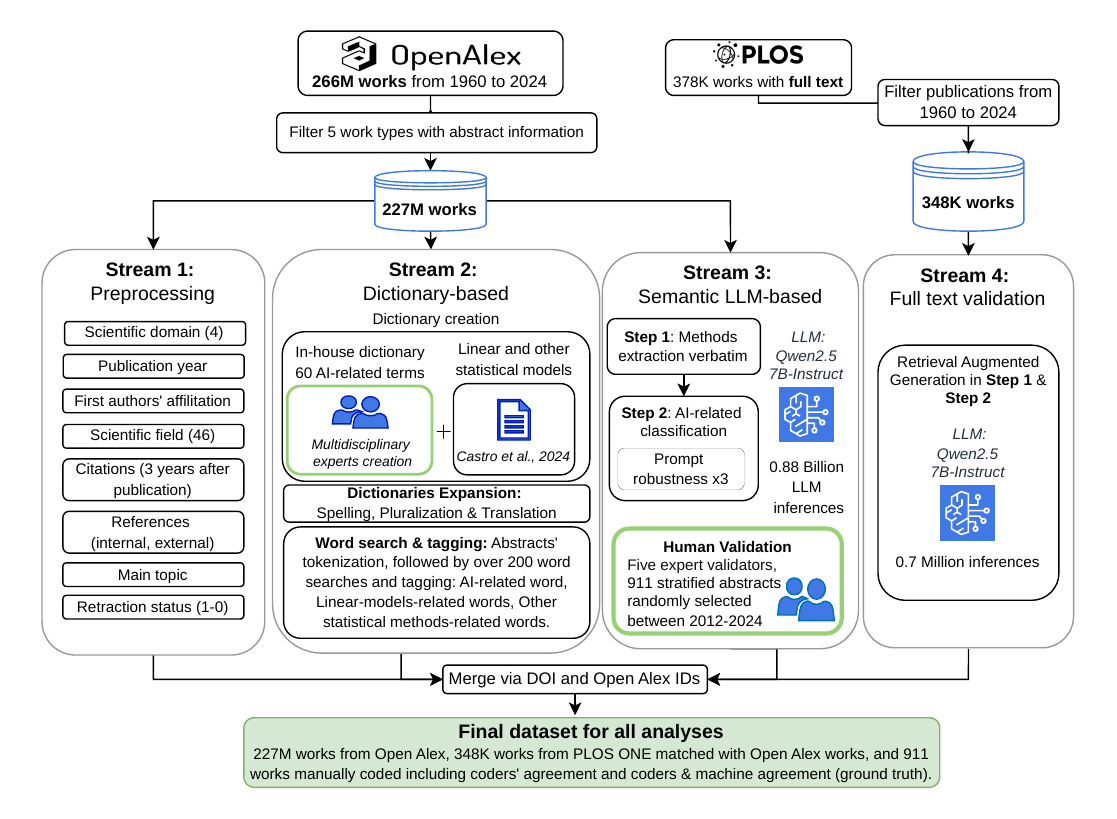}
  \caption{Flow chart of data processing and analysis streams based on the Open Alex collection and the PLOS ONE database of full text. Stream 1 describes the main bibliometric variables used in the analysis. Stream 2 describes the dictionary-based approaches, including the in-house-developed dictionary, the dictionary from \cite{castrotorres_2024_use}, and strategies for dictionary expansion. Stream 3 describes the two steps of the semantic LLM-based methods classification approach together with the general characteristics of the sample and procedure for Human Validation of the LLM’s outputs. Stream 4 refers to the Retrieval Augmented Generation procedure applied to the full text. All these streams are merged into a final data set which is the basis of all analysis.}
  \label{fig:feature}
\end{figure}

Our initial corpus comprises about 266 million works included in Open Alex database as of February 27th, 2025. We focus on five work types: articles, books, book-chapters, dissertations, and preprints with publication dates from January 1st, 1960 to December 31st, 2024. We excluded works with fewer than 200 characters in the abstract variable. The work-type and abstract-length criteria yielded a sample of 227,854,643 works. In this section, we present the architecture of the data pipeline used to process the sample. A summary of the architecture is shown in Figure \ref{fig:method}.

In \textit{Stream 1} we clean and create variables related to works’ field, citations in a three-year window after publication, referenced works classified in three categories (from the same field, from other fields, from Computer Science), topics, and the retraction status coded as one if the work has been retracted and zero otherwise. 

In \textit{Stream 2} (Dictionary-based), we used word search methods to capture the reporting of research methods in abstracts based on two lists of terms (hereafter referred to as “dictionaries”). Our first dictionary was built in our research laboratory (name to be disclosed later to preserve anonymity) through a two-step process. First, we provided a list of 94 terms to four colleagues with backgrounds in sociology, archeology, law, and media studies. Two of these colleagues were PhD students, and the other two were postdoctoral researchers with seven and five years of research experience. We asked them to classify each term as AI-related (1), potentially AI-related (2), non-AI-related (3), or unknown (4). We also asked them to add terms they considered AI-related that were not included in the initial list. After collecting their report and deduplicating their responses, we repeated the classification step. Following this iteration, our list comprised 115 terms, of which 60 were classified as AI-related by all four raters. We retained these 60 terms as our core dictionary of AI-related methods. The second dictionary was taken from \cite{castrotorres_2024_use}, who constructed it to capture the reporting of linear models (153 terms, e.g., logistic regression, ordinary least squares, generalized linear models) and other statistical methods (48 terms, e.g., principal component analysis, cluster analysis). To improve coverage, all dictionaries were expanded to account for common plural forms, British-English spelling, and naming variations (e.g., “logistic regressions” vs. “logistic regression”). We also translated methods’ names using automatic translation from the translate R package. Full-term lists and preprocessing details are provided in the Supplementary Materials.

In \textit{Stream 3} (Semantic LLM-based), we implemented a two-step pipeline using a large language model (Qwen-2.5-7B-Instruct) deployed on a local high-performance computing infrastructure. In the first step, the abstract is provided as input to the LLM, which extracts verbatim the sentences describing the study’s methodology. In the second step, the extracted sentences are fed back to the model, which generates a structured list of methods and classifies each as AI-related or non-AI-related. This two-stage approach reduces task complexity for the model and minimizes false positives from sentences where methods are mentioned incidentally, illustratively, or in commentary articles. The prompt templates used for each stage are shown in the Supplementary Materials. In total, we performed 0.88 billion inferences for this method, computing time was approximately 92 hours of compute across 96 NVIDIA A100 GPUs divided in 24 computing nodes of the HPC MareNostrum 5 located in Barcelona.

\subsection{Validation methods}

To validate the LLM-based method, we implemented a series of robustness checks. First, we tested prompt robustness by generating semantically equivalent variants and comparing outputs across three different formulations. Second, we manually annotated 911 randomly selected abstracts, each annotated independently by three people. These human annotations were followed by two steps for achieving agreement (1) among coders and (2) between the coders and the LLM. We use these agreement results to estimate the model’s precision, recall, and the F1 statistic, and we rely on Cohen’s Kappa and Krippendorff's Alpha to assess inter-coder and coder-machine agreement levels. Third, we applied an analogous prompting strategy to a sample of 372,323 full-text articles from the PLOS ONE database, which yielded results consistent with those from abstracts

\subsubsection{Prompt robustness} To evaluate the stability of the LLM-based method with respect to prompt formulation, we use three prompting strategies. The first was a baseline strategy where no answer examples were provided in the prompt template. The second strategy included fixed examples, in which eight AI-related and eight non-AI-related terms drawn from the dictionary-based method were provided as references. Finally, the third strategy introduced random examples, where 16 method examples (eight AI and eight non-AI) were randomly selected from the dictionary for each abstract classification.

Comparing outputs across these strategies allows us to test whether results hinge on specific prompt wordings or example choices. We anticipated that the fixed-example strategy might yield a higher proportion of empty outputs, reflecting a “downward bias” toward the provided examples, whereas the random-example strategy could increase positive responses by broadening the scope of terms shown. The baseline prompt, in turn, was expected to be unbiased but potentially less informative. This robustness check reduces sensitivity to prompt-specific artifacts and supports reproducibility of the findings. 

\subsubsection{Human Validation} 
To assess the validity of the LLM-based extractions, 911 abstracts were coded by five independent coders. Each abstract was randomly assigned to three coders. The sample size was determined based on human resources’ availability after a pilot coding test performed by two undergraduate students and two of our lab researchers on slightly over 100 abstracts. This pilot showed that, on average, undergraduate students needed twice the time compared to the researchers to detect methods in abstracts and classify them as AI- and non-AI-related (1.1 min vs 0.55 min). Additionally, the accuracy of undergraduate students was lower compared to that of experienced researchers. In light of these results, we decided to employ four researchers and one undergraduate student with a time constraint of up to six hours per person. Assuming that, on average, these persons will spend 40 seconds per abstract, five hours and a half of work per person translates into 5 persons x 60 min/hour x 5.5 hours x 1.5 abstracts/minute = 2,475 manually coded abstracts. To warrant that each abstract is coded by at least three people, our target sample size was calculated as 2,475 / 3 = 825. We added a 10.5\% extra to account for potential low-quality abstracts and abstracts in languages that coders cannot read for a total of 825 * 1.105 = 911 works.

The sampling frame included articles published from 2012 onwards, the year where a sustained use of AI is detectable in all domains. This reference period is therefore crucial to our conclusion of the exponential growth of AI adoption. We assigned sample probabilities that yield approximately the same number of academic works across three variables: the four scientific domains, articles published until 2018 and after, and the three categories of the LLM-based baseline classification: “No methods”, “Non-AI methods”, and “AI-methods.” This sample composition enhances our ability to equally assess accuracy, precision, and recall in the two targeted tasks across all scientific domains, while keeping focus on methods classification.

All coders participated in a 30-minute training session where instructions for methods identification and classification were explained. Once the coders went over their assigned abstracts, we conducted a between-coder agreement stage, followed by a coders-machine agreement. This latter step yielded our ground truth. The instructions provided to the coders are included in the Supplementary Material.

\textbf{Human Validation Results.} For methods identification, the model achieved a precision of 90.5\%, recall of 99.1\%, to an F1 score of 94.6 compared to agreed labels among coders. These figures indicate a high model’s reliability in detecting methodological mentions compared to what researchers agreed on. These figures increase to 92.4\% (precision), 99.6\% (recall), and 95.9 (F1) when LLM outputs are compared against the ground truth. Improved metrics further signal the suitability of the LLM approach for this task. 

Cohen's Kappa and Krippendorff's Alpha statistics at 0.58 reflected the complexity of identifying methods in abstracts. Values below 0.6 in these indicators suggest lack of strong agreement between the coders and the LLM (Cohen's Kappa), and among coders (Krippendorff's Alpha). Beyond the difficulty of the task, these relatively low agreement levels may be related to the diverse academic background of the coders including engineering (two coders), social sciences (two coders), and the life sciences (one coder) and their varying degrees of formal training: university (one coder), masters (two coders), and PhD level (two coders). Cohen's Kappa for coders against grounded truth is slightly higher at 0.66 further supporting the validity of our two-step approach for ground truth generation.

For AI methods classification, the model’s performance was slightly lower, and it improved when we use the ground truth as a benchmark vs only coders agreement. We indicate metrics against the ground truth in parenthesis. Thus, for the task of AI classification we estimate model’s precision at 83.0\% (85.8\%), recall at 96.2\% (97.9\%), and F1 score of 89.1 (91.4), which leads us to validate our approach. Importantly, the percentage of False Negatives in the AI methods classification was relatively low at 6.7\%. A figure that is seven times larger than the percentage of False Positives (0.88\%) which undermines the likelihood of overestimating trends of AI adoption in our analysis. Cohen's Kappa statistic suggests strong agreement between coders and the LLM at 0.81, which further improves when comparing the LLM with the ground truth 0.85. The Krippendorff's Alpha suggests a strong, yet imperfect, level of intercoders agreement with 0.64. 

Qualitative analysis of these errors highlighted predictable patterns: reviews or meta-studies discussing AI applications without actually employing AI, mentions of AI-related acronyms or terms not strictly referring to AI (e.g., Autoinducer, Artificial limb, Artificial Noise), and non-article texts such as special issue reports introduced some noise. In addition, the task proved challenging even for human annotators, with inter-annotator agreement of 79.1\% in the methods identification, and 75.4\% in the classification of methods into AI and non-AI. Difficult cases included domain-specific practices where methods are implied rather than explicitly reported (e.g., chemical studies), large meta-analyses that mention AI without applying it, and AI approaches that are not based on learning from data but exhibit adaptive behavior (e.g., Artificial Bee Colony, elastic net regularization). Despite these challenges, the high recall and strong overall performance indicate that LLMs are well-suited for systematic labeling of methods and AI techniques in scientific literature.

\subsubsection{Full-text extraction comparison} 
Analyses based on abstracts may underperform in cases where abstracts do not explicitly describe the research methods, methods’ descriptions deviate from conventional phrasing, or there are no conventions to refer to methods. This lack of conventional phrasing is likely the case of AI applications, particularly for more recent years. As AI applications for data collection and analysis become more frequent, we expect a reduced number of explicit mentions to AI methods in abstracts. To assess the extent of these potential omissions and their influence on our findings, we conducted a robustness check using a sample of 387,711 full-text articles obtained from PLOS ONE.

After excluding publications from 2025 onward, the full-text database contained 347,522 records published between 2003 and 2024. We first parsed the full text of each scientific paper, segmenting it into its constituent sections. To identify the methods section, we employed a keyword-matching approach, scanning section titles for terms such as "methods", "materials", "experimental", and "procedure". When no match was found, we fell back to a Retrieval-Augmented Generation (RAG) system that uses the all-MiniLM-L6-v2 sentence-transformers model to retrieve the sections most semantically similar to a methods-related query (Reimers and Gurevych 2019). The retrieved methods section was then incorporated into the prompt for classification, analogously to how we treated the abstracts. Keyword matching successfully identified the methods section in 95.5\% of cases, while only 4.5\% required the semantic similarity fallback.

To compare consistency between the two approaches (i.e., full text vs abstract), we merged the full-text results with those obtained from processing Open Alex abstracts. Overall, working with abstracts seems appropriate to extract general information on the usage of methods, however, this approach underestimates the use of AI. This potential underestimation is unlikely to affect time trends as we discuss below.

In terms of methods extraction, the abstract- and full-text-based approaches coincide in 94.5\% of cases, with 93.1\% of the cases retrieving methods’ related sentences according to the two approaches and only 1.4\% of cases retrieving no methods (4,868 works). There were 14,716  cases (4.3\%) where the abstract-based approach signaled the reporting of methods and the full-text did not, and 4,285 cases (1.2\%) where the reverse was true. The higher prevalence of methods detection in abstracts compared to full texts may be related to the structured nature of abstracts which can make it more straightforward to extract methods related to sentences and less prone to error.

As for methods’ classification, the overall prevalence of AI-related methods is higher according to the full-text analysis than in the abstract-based analysis. With the full text approach, we identify 26,390 AI-based works (7.6\%) vs. 7,202 (2.1\%) when we only use abstracts. This difference is expected given the richer methodological detail present in full texts. This detection gap varies substantially across scientific domains: 7.1\% vs. 1.6\% in the Life Sciences, 11.2\% vs. 2.4\% in the Social Sciences and Humanities, 17.0\% vs 5.0\% in the Physical Sciences and Engineering, and 4.5\% and 1.5\% in the Medical Sciences. 

The magnitudes of these gaps indicate that our main analysis represents a conservative estimate of the true levels of AI adoption in science. AI adoption is likely more prevalent than we estimate in Figures 2, 3 and 4. However, we doubt AI adoption in general is as high as estimated by the full text analysis in PLOS ONE data. These latter results are likely biased upwards due to the empirically- and quantitatively oriented nature of PLOS ONE publications. In addition, full text analysis is also prone to false positives, a feature we did not assess in this work.

Importantly, time trends in AI adoption between full text and abstract-based results are consistent (refer to Figure 3S in the SI), meaning that our conclusions regarding change over time hold. For example, up to 2019 the percentage of academic works reporting AI methods was 5.6\% and 1.1\% according to the full-text and abstract-based analyses, respectively. These two figures increased to 12.4\% and 4.6\% for the 2020-2024 period, signaling growth in both approaches. The fact that results based on abstracts display faster proportional growth than full-text results indicates that greater standardization in methods reporting probably goes hand in hand with increased AI adoption making it easier and more straightforward to report AI methods in abstracts in a distinguishable manner.

\section{Results} \label{sec:results}

\subsection{Heterogeneous Patterns of AI adoption across Scientific Domains}

The black bold lines in Figure 3 show the time trends in the percentage of academic works using AI methods as part of their research strategy (AI adoption) as a fraction of articles using any research method for the four Open Alex scientific domains. To obtain them, we input the works’ abstracts into an LLM and instructed it to extract and classify the reported methods using three variations. The baseline instruction included no examples (thick line), and the following ones included fixed (dashed line), and randomly chosen examples (dotted line) of AI and non-AI methods, such as machine learning and logistic regression, respectively. We refer to these works as AI-based works or AI-based research. The time trends of AI-based research for scientific fields are displayed in light grey and are based on the baseline instruction without examples. The red dashed line shows the trends for the share of academic works using any of 60 AI-related terms also as a fraction of articles reporting methods. This is our proxy for AI engagement as studied in the cited articles. The difference between AI adoption and AI engagement gives us a proxy of AI discussions, i.e., of works that use AI terminology but do not adopt AI technologies (refer to Figure 1S). 

For the Physical Sciences and Engineering, the higher prevalence of AI engagement over AI adoption may be due to works that develop AI methods, applications, and technologies and therefore name them in their abstracts without necessarily applying them to a research question. For the Social Sciences and Humanities, higher AI engagement potentially comes from works that comment, discuss, or problematize the adoption of AI methods and therefore also mention them in abstracts, albeit for different reasons. The fact that AI engagement is persistently over AI adoption in these two domains, suggests that methods development and discussions persist as AI become more widespreadly adopted. 

In contrast, there is no such AI-engaged literature in the Life Sciences and Health Sciences domains which presumably aligns with their disciplinary orientations. There is no scientific field within these two domains that has the development or discussion of research methods, notably AI methods, at the core of its research topics. As a result, AI engagement and AI adoption hovered around each other up until 2002 in the Life Sciences and 2014 in the Medical Sciences. It is only after these two years that the red lines appear consistently over the black one potentially indicating the emergence of AI discussions on the top of AI adoption.

\begin{figure}
  \centering
  \includegraphics[width=0.9\linewidth]{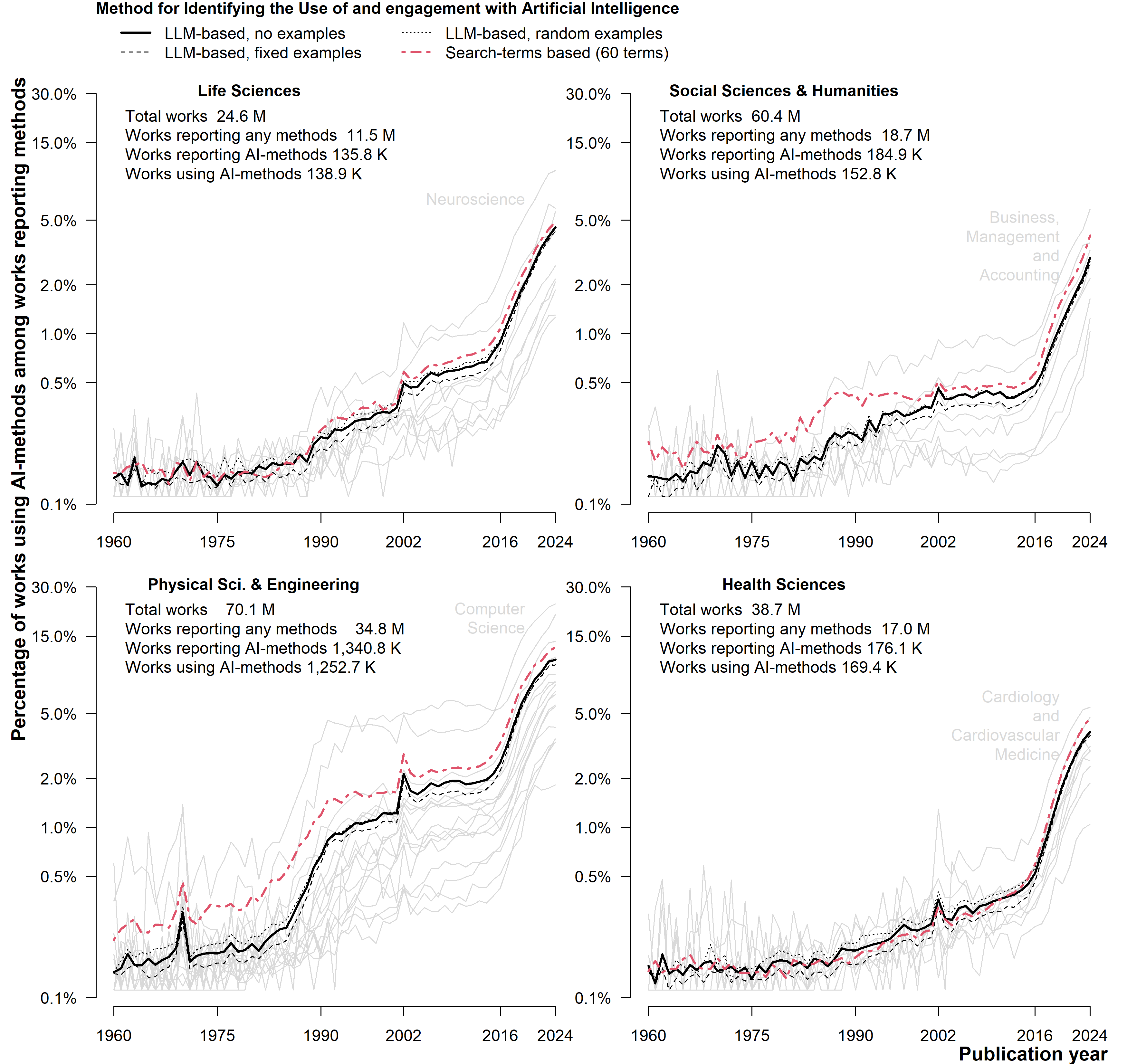}
  \caption{Log-scaled yearly percentage of academic works using AI methods (black and grey lines) and engaged with AI (red-dashed line) as a fraction of works using any method from 1960 to 2024 by scientific domains (bold lines) and 46 nested scientific fields and subfields (gray lines). Source: Authors’ calculation based on the Open Alex collection (n = 227,854,643 works, Data downloaded on Feb 27th, 2025).}
  \label{fig:feature}
\end{figure}

In terms of AI adoption, it was during the late 1980s when signs of sustained growth were observed in the Physical Sciences and Engineering, with Computer Science being the leading discipline. This sustained growth was preceded by a peak in the late 1960 consistent with descriptions of the pre 1990-AI developments as successive AI winters and springs \cite{mitchell_2020}. This result, consistent across all the three prompting strategies, means that AI developments translated into research applications a few decades after the field of AI was founded in 1950s and 1960s, and occurred primarily within the domain of the Physical Sciences and Engineering. By the 1990s, about one percent of the academic works reporting methods in this scientific domain relied on AI-based research methods. A sustained growth in this percentage followed this decade, certainly driven by the development of convolutional and recurrent neural network applications \cite{hochreiter_1997_lstm, krizhevsky_2017_imagenet, lecun_1998_gradient}, the introduction of word2vec and transformer architectures \cite{mikolov_2013_efficient, vaswani_2017_attention}, the reduced cost for training deep learning models thanks to Graphic Processing Units \cite{raina_2009_largescale}, and the subsequent proliferation of open-source frameworks for natural language processing (\cite{jurafsky_2009_speech}). Finally, around 2015, AI adoption exploded, reaching a prevalence of above 10 percent by 2024, i.e., more than one in every 10 published works relied on AI methods. Given that our analysis rely on abstract, these percentages are likely to be underestimated (refer to Data and Methods for further details) 

In scientific domains other than the Physical Sciences and Engineering, AI adoption has grown slower with a sustained linear growth (in the log scale) from the 1990s onwards. It is only around 2016 that AI adoption took off in these domains. Due to these delayed trends, the levels of AI adoption by 2024 are more heterogeneous among fields and far lower than those observed for the Physical Sciences and Engineering (10,7 percent): 4.4 percent in Life Sciences, 2,8 per cent in the Social Sciences and Humanities, and 3,8 percent in Health Sciences. A commonality across all domains is, however, that a few disciplines excel in AI adoption: Neuroscience in the Life Sciences (10 per cent in 2024), Business Management, and Accounting in the Social Sciences and Humanities (5,7 percent in 2024), Computer Science in Physical Sciences and Engineering (23,5 percent in 2024), and Cardiology and Cardiovascular Medicine in the Health Sciences (5,4 percent in 2024).

\subsection{Generalized growth, heterogeneous timing, and the critical role of Computer Sciences for AI adoption across scientific fields }

A further examination of field-specific trends confirms the dramatic pace of AI adoption in the last few years and reveals substantial field-specific heterogeneity in terms of the timing and scope of this trend. Figure 4 compares the yearly series of AI adoption for 46 scientific fields against their corresponding domain from 1985 to 2024. The thick circles in the series indicate the point in time after which AI adoption trends displayed five consecutive years of positive change after 2004. The y-axes are left free to favor the visualization of growth trends. 

According to this figure, the sustained growth of AI adoption started later for most scientific fields than their respective domains, meaning that the increasing trend of AI adoption in the mid-2000s resulted from a generalized but gradual adoption of AI in most scientific fields along with leading more prevalent applications in a few others. This pattern of generalized and pace-divergent growth across scientific fields fundamentally changed in the last few years where time series of AI adoption display exponential growth trends in all but one field: Mathematics, and to a lesser extent in Energy. And yet, even in these two fields, the growth in AI adoption measured with respect to the year 2005 has been at least fourfold.

More generally, all scientific fields, including the group of works with missing field information, display at least a fourfold increase in AI adoption from 2005 to 2024, implying annual growth rates of at least 7.3\%. These large growth rates are found across scientific fields of all sizes from below one-million-work fields such as Veterinary (6.8 growth in AI adoption) and Dentistry  (23 growth in AI adoption), to scientific fields that comprise over seven million works such as Arts and Humanities (12 growth in AI adoption) and Medicine (20 growth in AI adoption). The only exception to the over fourfold growth pattern is the field of Control and Systems Engineering which displays an almost threefold increase in AI adoption from 2005 to 2024 (i.e., 2.9 growth in AI adoption). Importantly, this is the second field with the largest proportion and the earliest adoption of AI, only surpassed by Computer Science. 

Notably, 17 out of the 46 fields (i.e., 36\% of the fields) in Figure 4 display over tenfold growths in AI adoption from 2005 to 2024, i.e., over 10\% annual growth rates. Four of these fields are in the Life Sciences, four in the Social Sciences and Humanities, only one in Physical Sciences and Engineering, and eight in the Health Sciences speaking to the significance and spread of AI adoption, and indicating that growing trends are likely to continue in the years to come; there is no evident sign of stagnation in AI-based research in any of these fields.

\begin{figure}
  \centering
  \includegraphics[width=0.9\linewidth]{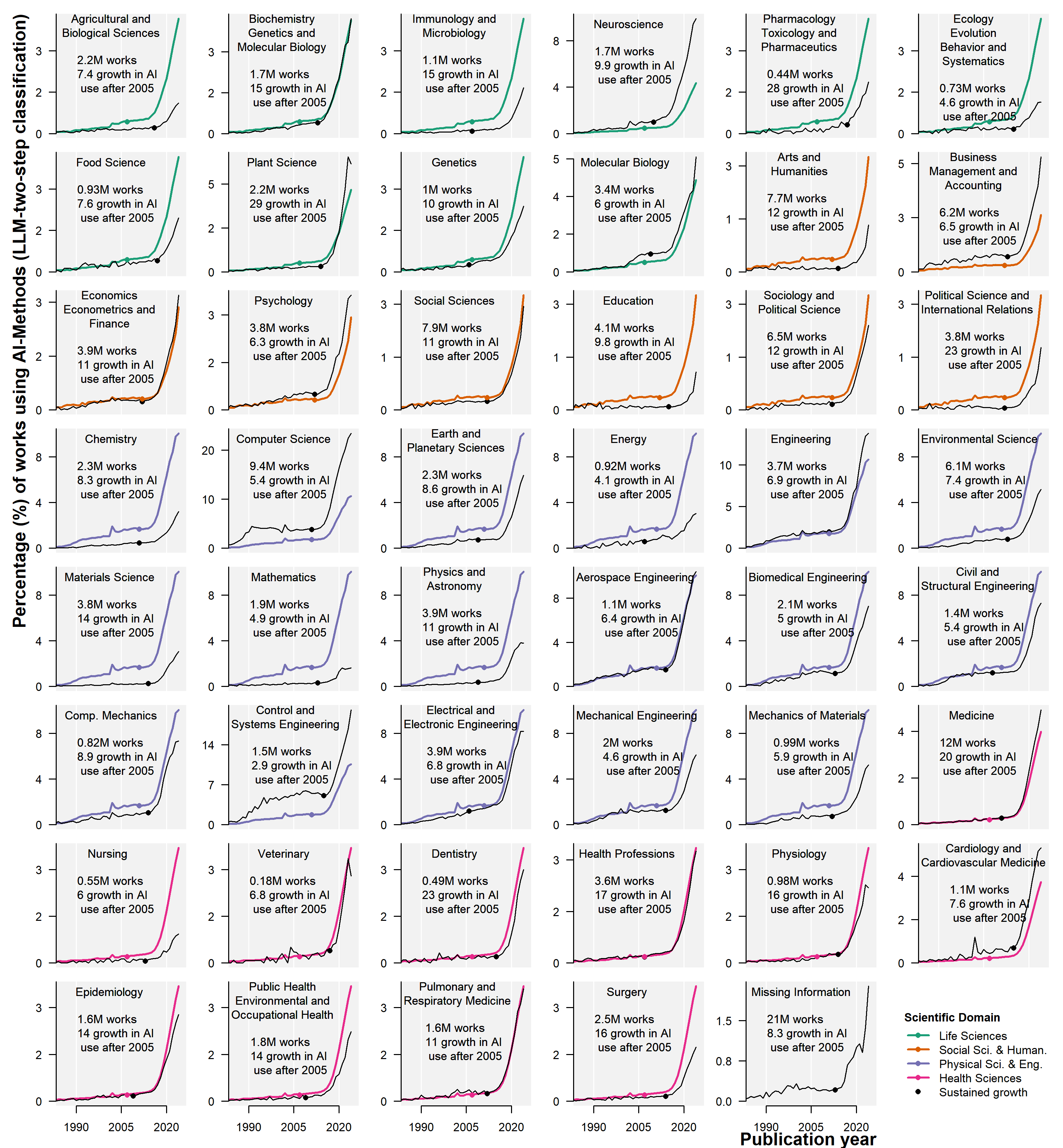}
  \caption{Yearly percentage of scientific works using AI methods as a fraction of works using any method for four scientific domains (colored lines), 46 scientific fields and subfields (black lines), and for works with missing scientific domain information from 1985 to 2024. The total number of academic works per field and the growth of AI use refer to the period 2005 to 2024 (n = 143,232,964 works).}
  \label{fig:feature}
\end{figure}

This generalized growth has led to unequal levels of AI adoption across fields. In Life Sciences, three fields excel with above five per cent levels of AI adoption in 2024: Neuroscience, Molecular Biology, and Plant Science; all other fields display much lower percentages including values below one per cent in Agricultural and Biological Sciences, and Ecology, Evolution Behaviour and Systematics. Likewise, in the Social Sciences and Humanities, the field of Business, Management, and Accounting excels with over five per cent articles using AI in 2024, and contrast with the below-one-percent AI adoption in Education, and Arts and Humanities in the same year. In the Physical Sciences and Engineering, the lowest levels of AI adoption in 2024 are observed in the fields of Chemistry, and Material Sciences; yet these two fields are above most scientific fields in other domains. The two leading fields in this domain are Control and Systems Engineering, and Computer Science. Finally, in the domain of Health Sciences, AI adoption in the last year of analysis is relatively more homogeneous with levels hovering between three percent and four percent. The only exception is the already mentioned leading field of Cardiology and Cardiovascular Medicine.

The patterns discussed until this point put in evidence the critical role of Physical Science and Engineering in leading the adoption of AI in research since the 1990s, particularly through the fields of Computer Science and Control and Systems Engineering. The field of Computer Science excels by the early rising trends on AI adoption. Other scientific fields and domains display patterns of delayed adoption with clearly differentiated leading fields. These field-lead trends lasted a bit more than 10 years from the 2000s to 2016. The explosion of AI adoption after 2016 is something quite different spreading across virtually all fields of science in an exponential fashion.

\subsection{Differential Pre- and Post-publication Patterns for AI-based Research }

The accelerated trends of AI adoption have been accompanied with very apparent pre- and post-publication patterns, all of which signal untapped epistemological implications due to AI adoption. Specifically, compared to works that use non-AI methods, AI-based works published between 2004 and 2024 tend to concentrate on a fewer topics; rely more extensively on existing analysis frameworks in quantitative research; be more interdisciplinary, as per citations to Computer Science works, but not to citation to other fields; be more likely to be cited at least once; and display higher retraction rates upon publication. The five panels in Figure 5 support these claims. 

\begin{figure}
  \centering
  \includegraphics[width=0.9\linewidth]{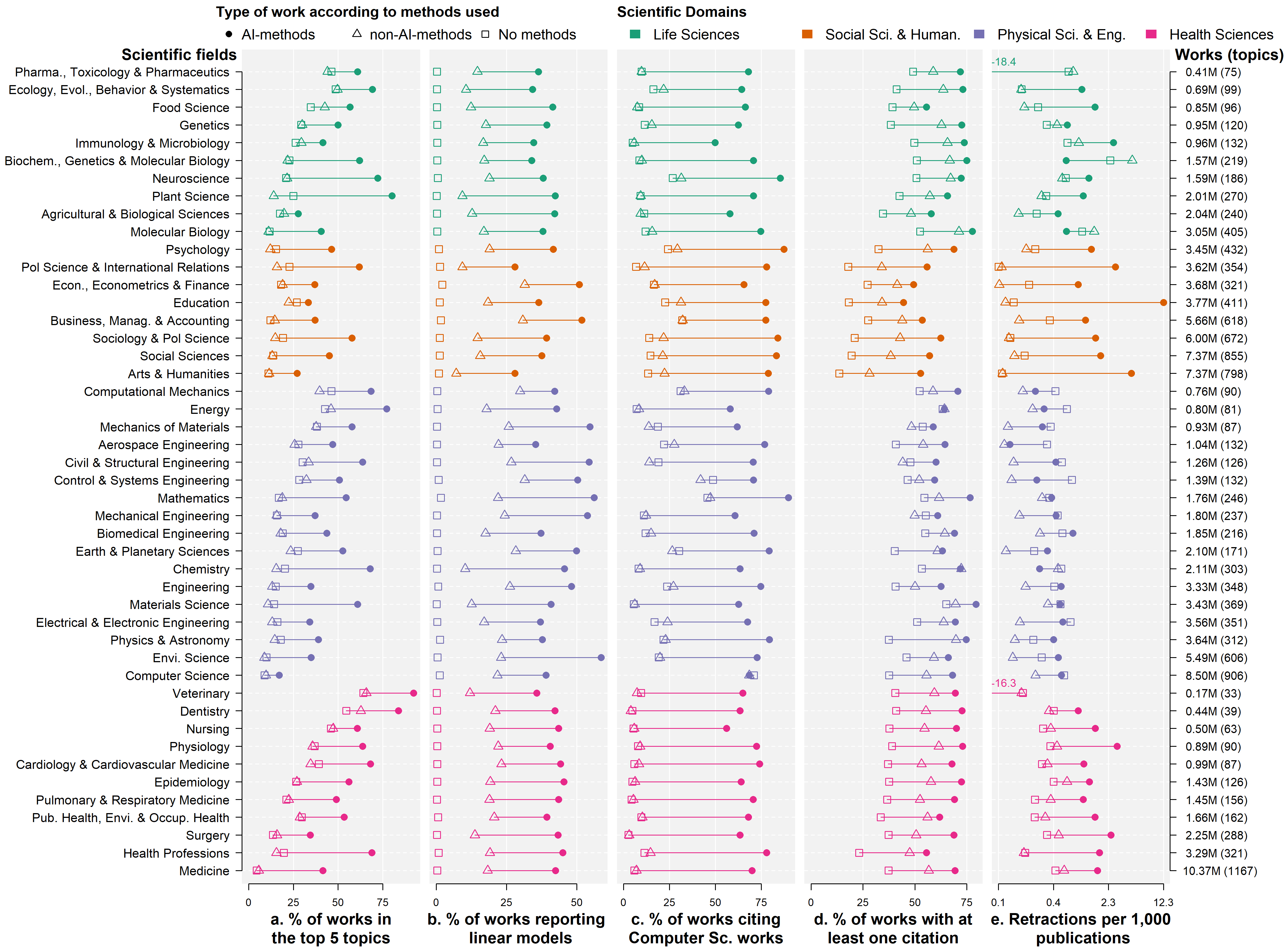}
  \caption{Pre- and post-publication patterns for academic works according to the type of methods: Artificial Intelligence methods (AI-methods), non-AI-methods, and No methods. The use of methods is determined by the baseline prompting strategy without examples (refer to Materials and Methods sections). Panel a. reports raw percentages. Panels b. to e. report predicted percentages and rates from a multivariate quasi-Poisson model that control for: publication type ("article", "book" , "dissertation", and "preprint"), publication year ("[2000,2005)", "[2005,2010)", "[2010,2015)", "[2015,2020)", "[2020,2025]"), and language ("English", "German",  "Spanish", "French", "Indonesian", "Portuguese", "Other"). Predictions correspond to categories underlined in italic. The analytical sample for all panels except panel (c.) includes 143,232,964 works published between 2002 and 2022. The analytical sample for panel (c.) excludes articles with missing information on cited works (25,994).}
  \label{fig:feature}
\end{figure}

\textbf{Topic diversity.} AI adoption is characterized by restricted applications, topic-wise. According to Panel A, AI technologies are being used in a few topics where they fit, meaning that the opportunities for expansion are largely untapped. The panel (a) in Figure 5 displays the percentage of articles in the top five topics for each of the method usage groups: AI-methods, non-AI-Methods, and No methods. High percentages indicate a concentration of research works in  the top five topics, whereas lower values indicate increased topic diversity. Despite the heterogeneity in the number of topics across fields, topic diversity seems generally lower among AI-based works. In most scientific fields, the share of works in the top five topics is 20 percentage points higher among works using AI compared to others, whereas there is no apparent difference in topic diversity between works reporting non-AI methods and works without methods. A handful of fields display smaller gaps: Agricultural and Biological Sciences in the Life Sciences, Education in the Social Sciences and Humanities, and Computer Science in the Physical and Engineering Sciences, and to a lesser extent Nursing in the Health Sciences. Yet, in all these cases, the gap in topic diversity follows the same direction. Panel (a) in Figure 2S shows that the pattern of lower topic diversity among AI-based work compared to the rest holds when using the Shannon Entropy measure on works’ distribution across topics, with lower standardized Shannon Entropy for AI-based research compared to non-AI-based research. 

\textbf{Analysis frameworks.} The disruptive nature of AI is also tapped by its reliance on existing analytical frameworks, at least in quantitative research. Panel (b) in Figure 5 displays the predicted proportion of articles reporting linear models in their abstracts as a proxy for works’ reliance on the linear model analytical frameworks \cite{lieberson_2002_barking, cornwell_2015_social, abbott_1988_transcending, gollac_2004_rigueur, castrotorres_2025_hegemonic}). Given the widespread use and flexibility of generalized linear models in quantitative research, these predicted percentages serve to assess whether AI adoption goes along with the use of widespread analytical frameworks or represents a methodological breakthrough. Results are more supportive of the former, which aligns with decreased topic diversity among AI-based works. 
Thus, not only is AI adoption limited to a few topics, but it also is limited in terms of analytical frameworks. In all scientific fields, academic works that use AI methods are more likely to report linear models compared to academic works that use non-AI methods. Between one-third and one-half of works using AI simultaneously report linear models (e.g., linear regression, logistic regression). These predicted percentages hover around 25 per cent for works reporting non-AI methods. The few exceptions to this rule include Economics, Econometrics, and Finance; Business, Management, and Accounting; Computational Mechanics; Control and System Engineering; and Earth and Planetary Science. In four out of these five fields, however, the predicted percentage of works reporting linear models among AI-based works surpasses 50 percent. These greater proportions of linear models reporting between works using AI and works using other methods signal a potential dependency of AI applications on existing analytical frameworks. 
The panel (b) in Figure 2S shows that statistical methods outside of the generalized linear model framework (proxied by the search of 48 terms) are less frequently used in all fields among both AI-based and non-AI-based research, which further support the notion of linear models’ hegemony, to use Castro and Akbaritabar’s (\cite{castrotorres_2025_hegemonic}) expression, and suggest that AI-based research is potentially part and parcel of this pattern.

\textbf{Citations to Computer Science works.} AI adoption is accompanied by a higher likelihood of citing research from Computer Science, signaling the key role of this field for widespread AI adoption patterns. Panel (c) in Figure 5 shows the predicted percentage of academic works citing at least one work from Computer Science. The pattern is very clear: Computer Science has been fundamental for AI adoption, as almost 75\% of AI-based work cites at least one paper from this field. These percentages are below 30\% for non-AI-based works and for research works reporting no methods, implying above-40 percentage-point citation gaps in most, if not all, fields. By default, these citation gaps do not exist within Computer Science. Panel (c) in Figure 2S shows, however, that out-of-field citations to other fields as predicted by the models (excluding citations to Computer Science) are equally prevalent among AI-based and non-AI-based research, except perhaps in a few fields within the Health Sciences where AI-based research displays a higher predicted percentage of out-of-field citations compared to non-AI-based works (e.g., Cardiology and Cardiovascular Medicine, Epidemiology, Pulmonary and Respiratory Medicine, and Surgery); these patterns require further research. In addition, these predicted out-of-field citations display levels only slightly above those of the citations to Computer Science among AI-based research (i.e., around 75\%), meaning that for AI-based research, the field of Computer Science alone has attracted as much scholarly attention as all fields combined for research in general, regardless of the methods used. 

\textbf{Works’ visibility.} In terms of post-publication patterns, Panel (d) shows that AI-based works display a higher likelihood of receiving at least one citation within the first three years after publication compared to other works. Remarkably, across all fields except Education, more than 50 per cent of academic works using AI methods received at least one citation within the first three years after publication. These citation gaps operate on top of differential citation levels across scientific domains, with lower proportions of cited works in the Social Sciences and Humanities compared to the other domains. Additionally, the percentage point gaps in works’ visibility between AI-methods and No methods works are remarkable, with above-20 percentage points differences for most fields. A notable single exception is the field of Energy, where the percentage of articles receiving at least one citation is above 60\% for all three types of work. Panel (d) in Figure 2S confirms the significance of these visibility gaps by comparing the predicted proportion of highly cited articles, i.e., works with at least seven citations. As in Figure 5, AI-based research displays the largest proportion of highly cited works with only a few exceptions in four fields in the Physical and Engineering Sciences.

\textbf{Retraction rates.} Finally, panel (e) displays the number of retracted academic works per 1,000 publications. The logarithmic scale is used to favor readability. As for results in panels b, c, and d, these predicted retraction rates account for publication type, language, and publication year. Thus, net of these works’ characteristics, academic works that use AI are consistently more likely to be retracted in all fields in the Social Sciences and Humanities; these retraction gaps by methods used are the largest of all scientific domains ranging from around one retraction per 10,000 publications (0.1 retraction rate) to 12.3 retractions per 1,000 published works in the field of Education. The Health Sciences domain follows closely these patterns, although with much smaller gaps and one exception, namely, Veterinary, where the predicted retraction gap among AI supported works is virtually zero (i.e., exp(-16.3) = 8,3 x 10-8). Except for this field, in all the rest, academic works that use AI methods display higher retraction rates than works that use other methods. These higher retraction rates are significant. Even in the case with the lowest gap, namely, Dentistry, they implied more than doubled retraction rates: 0.44 vs 0.92 retractions per 1,000 publications.

Among scientific fields in the Life Sciences and Physical Sciences and Engineering, results are mixed, with fields in the latter domains having, overall, the lowest retraction rates of all. In these two domains, retraction gaps among academic works according to the methods used are smaller compared to other domains, and they are not in the same direction. In the Life Sciences, six out of the 10 fields display larger predicted retraction rates among academic works using AI methods compared to the other two groups. The exceptions are Pharmacy, Toxicology and Pharmaceutics; Biochemistry, Genetics, and Molecular Biology; and Molecular Biology. In the Physical Sciences and Engineering, only four out of 17 fields have academic works using AI in the leading position in terms of retraction rates: Biomedical Engineering; Earth and Planetary Sciences; Engineering; and Physics and Astronomy. The retraction rates reported in the Panel (e) in Figure 2S depict similar patterns across fields along with larger gaps with AI-based research having higher retraction rates. The fact that this panel focuses on works with at least one citation indicates that increased visibility among AI-based research does not fully explain retractions.

Taken together, these patterns suggest immediate challenges for continuing AI adoption in scientific research. While retractions are a good sign for science systems, large gaps should also indicate that structural corrections may be needed. Moreover, these retraction rates patterns suggest a clear divide between fields in Physical Sciences and Engineering, the fields that simultaneously develop and use AI methods, and scientific fields and domains that are more likely users of AI methods. This is very clear in the case of the Social Sciences and Humanities, the domain with the largest retraction rates among AI-based works, and to a lesser extent across fields in Health Sciences also arguably an AI user, not a developer, and the Life Science. 

\begin{figure}[!hp]
  \centering
  \includegraphics[width=0.9\linewidth]{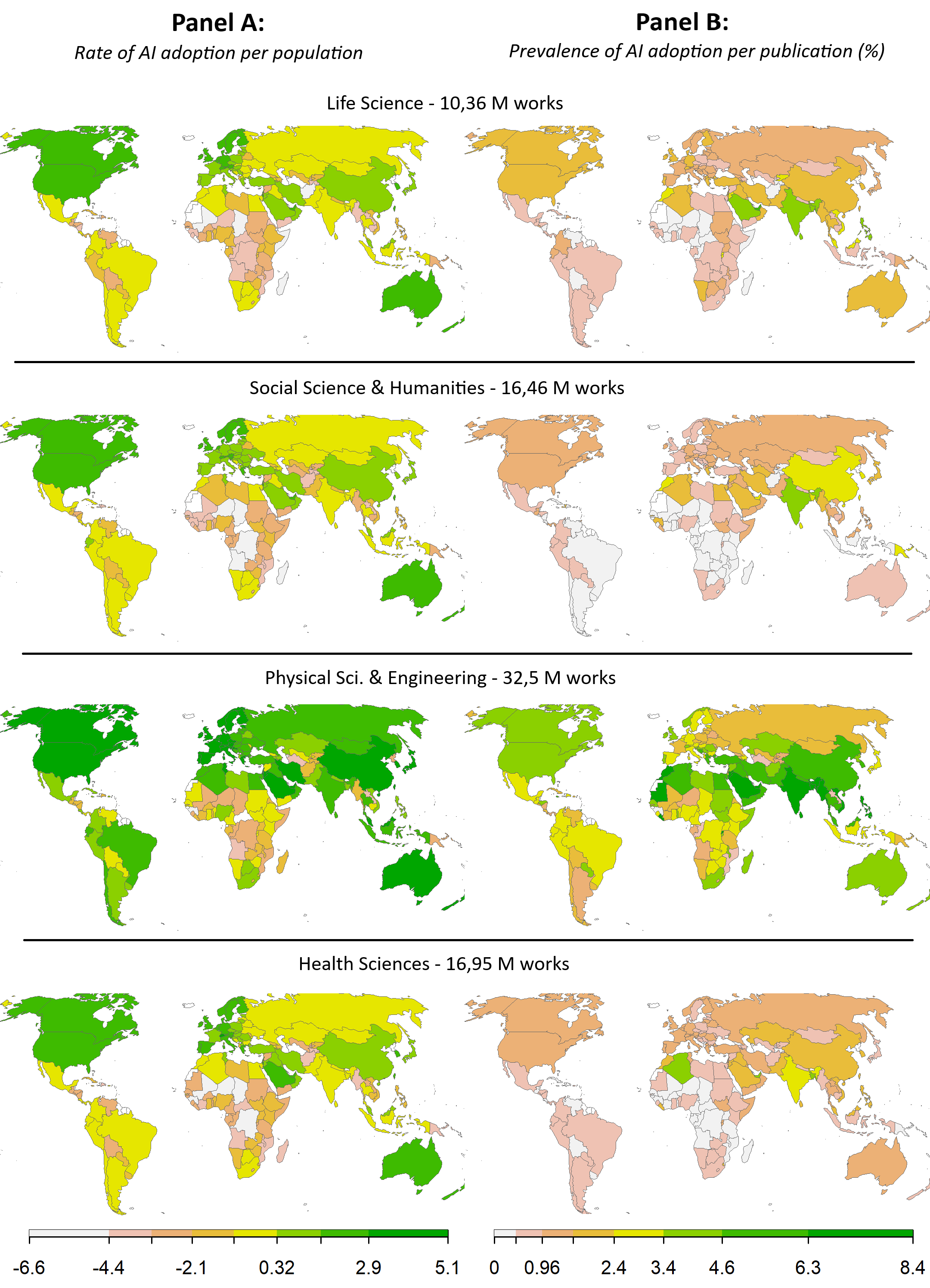}
  \caption{Country-level rates of AI adoption in science per 100,000 people (Panel A, logarithmic sale) and percentage of AI adoption per publication (Panel B) for academic works published between 2002 and 2024. We include information from first authors’ country of institutional affiliation and for countries with at least one million people and 100 publications from 2002 to 2024. This inclusion criteria comprises 77,472,370 academic works, that is, 99.9\% of all publications of the period (6,363 publications are excluded).}
  \label{fig:feature}
\end{figure}

\subsection{The Two Sides of the AI adoption Geography}

The geographical distribution of AI-based works adds a layer of complexity to the challenges of increased AI adoption. According to Figure 6, augmented diversity and persistent inequalities across countries are the apparent features of AI adoption geography.

Measured by the number of AI-based works by population (Panel A in Figure 6), the geographical patterns of AI adoption are consistent with existing studies on the spatial distribution of academic works \cite{montano_2026_transformation}: country-level AI adoption rates are the highest in countries with developed economies (i.e., North America, North and West Europe, Australia) along with emerging national research system in East Europe and Asia including countries such as Poland, China, Saudi Arabia, and Iran. The emergence of some global South countries is more apparent in the Physical Sciences and Engineering than in other domains where high rates of AI adoption per population spread beyond global North countries to virtually all world regions except Sub-Saharan Africa. The low levels of AI adoption in this latter region is the main evidence of persistent inequalities across global science systems.

The other side of AI’s geography further underlines the countries’ heterogeneous approach to AI adoption in science. Measured as the percentage of academic works using AI as a fraction of all published works (Panel B in Figure 6), the results suggest a swapped rank where the so-called emerging countries in terms of AI adoption rates, lead in terms of the prevalence of AI in a spatial belt that goes from Indonesia to Algeria. The North-South divide in this measure suggests that some countries in the global South are focusing resources on AI adoption in sciences, whereas research in the global North (perhaps due to its sheer volume) tends to remain more diverse method-wise. The case of India and neighboring countries is significant as they excel in AI adoption prevalence while not ranking at the top in the AI adoption rates.

\section{Conclusions and discussion}

The massive scope and open-access nature of the Open Alex collection enabled us to conduct a historical, global, and granular study of AI adoption across scientific domains and fields. If the 1960 to 1990 period can be characterized in terms of subsequent AI winters and springs, the 1990 to 2010 period seems like a continued spring in Computer Sciences and Control Systems Engineering with peaks associated with turning technological points, spreading developments, and bridging applications to other scientific domains and fields. The trends for the last ten years show an indisputable exponential growth of AI adoption with several implications.

These spin-off effects flourished in the early 2000s, triggering AI adoption in scientific research with clear exponential and generalized growth patterns after 2015 in all but two fields: Mathematics and Energy. The resulting levels of AI adoption in 2024 are very heterogeneous across scientific fields, and generally below 5 percent with domain-specific exceptional leading fields such as Neuroscience in the Life Sciences; Business, Management and Accounting in the Social Sciences and Humanities; Computer Sciences, and Control and System Engineering in the Physical and Engineering Sciences; and Cardiology and Cardiovascular Medicine in the Health Sciences. 

The specificities of AI-based publications compared to non-AI-based works in terms of pre- and post-publication patterns underline the challenges and opportunities ahead of growing AI adoption. Decreased topic diversity, greater reliance on linear model analysis frameworks, and a higher likelihood of citations to Computer Science works (but not to other fields) compared to non-AI-based research suggest that AI adoption’s potential is still untapped across most scientific fields. To the extent that topical, methodological and disciplinary diversity in research are positive to science, continued attention to how AI adoption evolves is important to avoid unwarranted or hype-driven patterns of scholarly attention and research funding concentration.

The post-publication patterns of AI-based research, namely, greater visibility in the academic community and greater retractions rates, further highlight the challenges ahead of continued AI adoption in science. As far as scientific production is a social endeavor, research institutions, universities, journals, and individual scholars should not see themselves as immune to the influence of the AI hype \cite{guest_2025_uncritical}. Previous research has shown that research methods and practices (including citation practices) do not necessarily obey inherent fitness or quality \cite{koppman_2019_methodological, leahey_2005_alphas, nielsen_2021_global, pereira_2024_rethinking}. Methodological choices and citations can be biased and therefore scholarly research must continue to use a critical lens on new methods and how they enhanced or harmed research practices. Although greater visibility, measured by citations, is positive, it is not fully clear that the large citation gaps between AI- and non-AI-based research are justified and necessarily beneficial for science. Adding larger retraction rates for AI-based research to this picture further underlines the need for greater attention to how, where, and who are the institutional leaders of AI adoption. 

Considering these opportunities and challenges, the principles of research openness, transparency, and ethics are crucial to track and harvest the benefits of increased AI adoption. Existing frameworks for data sharing, reproducibility, and research ethics provide the foundations to enact these principles. However, AI technologies, particularly those that rely on LLMs, include features that could make them hard to enact (Denton et al. 2021). For example, there are multiple layers to pre-trained LLMs starting from the origin and modes of gathering of the original data, preprocessing steps, models’ adjustments and ad hoc parametrization (e.g., the so-called guard-rails and sub prompting), and finally how users interacted with LLMs to produce results (e.g., prompting/querying strategies). Full openness, transparency, and ethics across all these layers are hard to enact and check, and the existing frameworks may be ill-prepared to deal with the multiplicity of use cases as AI technologies continue to develop, become more complex, and spread.

The increased geographical diversity of academic outputs, while being positive for science, adds a layer of complexity to AI adoption as it requires the principles of openness, transparency, and ethics to adapt and account for a variety of institutional arrangements across countries. Increased connectivity and international research collaboration will demand comprehensive and well-founded regulations for information and data sharing, which are heightened needs in several cases of AI-based research. Likewise, the challenges of helping countries with lower levels of AI adoption is fundamental to prevent larger knowledge production inequalities between global South and North countries, particularly with countries in the Sub-Saharan Africa region.

Finally, this is not the first time scientific research is widely impacted by technological change. The ongoing increase of quantitative data, statistical methods, and computational capacity from the 1960s onwards \cite{ruggles_2014_bigmicrodata, lieberson_2002_barking}, the spread of the Internet, and the incessant pressure from wars and societal crisis on scientific communities, have demonstrated the malleable character of science \cite{porter_1995_trust, harding_2015_objectivity}. Studies have also underlined science systems’ enduring characteristics such as disciplinary boundaries, persisting inequalities, and the hegemonic use of specific analytical frameworks. Greater attention to these historical accounts could help guide AI adoption to ensure that the principles of openness, transparency, and ethics are integral to the roots of the AI technology tree.

\begin{acknowledgments}
Andres Castro received financial support from the Spanish Ministry of Science, Innovation and Universities (Grant no. RyC RYC2023-042730-I by MICIU/AEI /10.13039/501100011033 y por el FSE+). Joan Giner-Miguelez acknowledges his AI4S fellowship within the “Generación D” initiative by Red.es, Ministerio para la Transformación Digital y de la Función Pública, for talent attraction (C005/24-ED CV1), funded by NextGenerationEU through PRTR. 
\end{acknowledgments}


\appendix
\pagebreak
\section*{Supplementary Material:}
\subsection*{1 - List of Terms for Dictionary Approaches:}

\textit{Artificial Intelligence-related terms}

"artificial neural network; artificial neural networks; computer vision; convolutional neural network; convolutional neural networks; deep learning; deep neural network; deep neural networks; image classification; image recognition; large language model; large language models; machine learning; natural language processing; neural network; neural networks; optical character recognition; perceptron; random forest; random forests; recurrent neural network; recurrent neural networks; reinforcement learning; self-learning model; self-learning models; supervised learning; support vector machine; training data; training model; training models; transformer model; transformer models; unsupervised learning; activation layer; activation layers; base tuning; text embedding; text embeddings; explainable ai; f1 score; federated learning; few-shot learning; zero-shot learning; fine tuning; genetic algorithms; instruct tuning; lstm network; lstm networks; mixture of experts; naive bayes; pre-trained embeddings; pre-trained models; quantum machine learning; quantumization; retrieval-augmented generation; reasoning models; retrieval interleaved generation; softmax; swarm intelligence; tf-idf; term frequency-inverse document frequency."\\

\textit{Linear models-terms (non-AI)}

"fixed effects model; fixed effects models; instrumental variable; randomized control trial; randomized control trials; regression discontinuity; analysis of variance; analyses of variance; anova; anovas; generalized linear model; generalized linear models; hierarchical model; hierarchical models; linear model; linear models; linear probability model; linear probability models; linear regression; linear regressions; logistic model; logistic models; logistic regression; logistic regressions; logit model; logit models; multilevel model; multilevel models; multilevel regression model; multilevel regression models; multinomial model; multinomial models; multinomial regression model; multinomial regression models; multivariate model; multivariate models; negative binomial model; negative binomial models; ordinary least squares regression; poisson model; a poisson model; poisson models; poisson regression; a poisson regression; poisson regressions; probit model; probit models; probit regression; probit regressions; regression analysis; regression analyses; regression model; regression models; survival analysis; survival analyses; time series model; time series models; zero-inflated negative binomial model; zero-inflated negative binomial models; zero-inflated poisson model; zero-inflated poisson models; accelerated time failure model; accelerated time failure models; age-period-cohort model; age-period-cohort models; arima model; arima models; autoregressive integrated moving average model; autoregressive integrated moving average models; cox model; cox models; cox regression; cox regressions; cox regression model; cox regression models; generalized additive model; generalized additive models; hazard model; hazard models; propensity score matching; proportional hazard model; proportional hazard models; random effects model; random effects models; sars model; sars models; species-area relationship model; species-area relationship models; ols; difference in difference; difference-in-difference; control for; controlling for; regression; log-linear model; log-linear models; log-log model; log-log models; manova; lasso regression; mixed effects model; multinomial logistic regression; multinomial logistic regressions; nonparametric regression; nonparametric regressions; ologit; ordered logit; quantile regression; quantile regressions; ridge regression; ridge regressions; robust regression; tobit; nested model; nested models; multi-level model; multi-level models; multi-level regression model; multi-level regression models; ancova."\\

\textit{Non-Linear models-terms}

"factorial analysis; factorial analyses; geometric data analysis; geometric data analyses; multivariate descriptive statistics; multivariate descriptive model; multivariate descriptive models; cluster analysis; cluster analyses; correspondence analysis; correspondence analyses; discriminant analysis; discriminant analyses; latent class analysis; latent class analyses; model-based cluster analysis; model-based cluster analyses; multiple correspondence analysis; multiple correspondence analyses; multiple factorial analysis; multiple factorial analyses; principal component analysis; principal component analyses; agent based modelling; network analysis; network analyses; sequence analysis; sequence analyses; simulation; simulations; gibbs sampler; gibbs sampling; markov chain; markov chains; markov chain monte carlo; optimal matching; agent-based modelling; qualitative comparative analysis; qualitative comparative analyses; agent-based modeling; agent based modeling; individual-based modeling; multi-agent-based modeling; individual-based modelling; multi-agent-based modelling; microsimulation; agent-based; multi-agent model; multi-agent models."

Translations are available upon request.
\pagebreak

\subsection*{2 - Instructions Provided to the Coders:}

There are about 550 abstracts in the sheet named “abstracts.” For each of them, we ask you to do the following:

\begin{enumerate}
    \item Read the abstract focusing on identifying sentences that describe the methods used in the article. By methods used we mean any kind of data collection or data analysis procedures that was used by the authors and that contribute to answering their research question directly or indirectly, or to make a case or push forward an argument. 
    \item Answer sequentially the following two questions according to your understanding of the abstract. 
    \begin{itemize}
        \item[Q1:] Are research methods reported in this abstract or can they be inferred reasonably?
        \item [Q2:] If yes, are these methods based on any kind of AI technology according to your own understanding of what AI technologies are? 
    \end{itemize}
Enter your responses in the corresponding columns using the following numerical conventions: 0 if you answer NO, 1 if you answer YES.
\end{enumerate}

\textbf{Important:} No other coding is available, so we are asking you to make your best guess.  If you like to think about it in probability terms, do so: how likely is it that this research used a method although it is not explicitly mentioned? How likely is it that the algorithm or abbreviation they mention (e.g. EGV genetic algorithm) is based on AI? If you feel that that likelihood is high, then answer YES. We understand this can lead to subjective judgement. That is fine. 
Important: Stick to your current knowledge; answer based on what you know. Do not go beyond the abstract content, except if you  need to translate a few words.

\textbf{Important:} Answer NO in Q1 when the quality or language of the abstract prevents you from a proper understanding of its content. Answer NO in Q2 if you do not know enough about the method, methodologies, or procedures mentioned in the abstract and you feel unable to classify them in a sensitive manner.

According to our own manual coding of 550 articles each, 2/3 of the cases should be straightforward, whereas 1/3 can be tricky. See below some hints that can help you with these latter cases.
\begin{itemize}
    \item Data can be qualitative, quantitative, large, small, numeric, textual, a collection of experiences or written texts (e.g., novels, biographies), self-participation, visual, memories, diary notes. Thus, methods can be any explicit or implicit reference to the procedures to capture, collect, or analyse any type of data. Analysis, description, synthesis, catalysis, experimentation, all these procedures indicate a method.
    \item There are no methods when the articles develop a theory or discuss concepts, or when they describe the content of an Special issue or summarize its articles. In most cases, when there are no data, there are no methods, and vice versa.
    \item Articles that report literary studies, meta-studies, systematic and non-systematic literature reviews count as articles having methods although they may be not explicitly mentioned as such. This recommendation is particularly relevant for articles from the Arts and Humanities. 
    \item Disciplines such as Medicine, Chemistry, Biology (especially Molecular Biology), and others often report experiments’ results without using the word experiment or without referring to the specific procedure that yielded the results (e.g., a chemical reaction, bacterial crops, etc.). We ask you to detect as much as possible the implicit method/procedure used in the research, and again, make your best guess.
    \item Note that you are labeling AI-related methods that may go beyond machine learning. Machine learning is a subfield of AI. Non-ML AI algorithms as evolutionary and genetic algorithm families are considered AI, despite not being directly machine learning.
\end{itemize}
\pagebreak

\subsection*{3 - Prompts used:}

In the first stage: 
\begin{researchpromptbox}

The following text is an abstract of a scientific work. Extract, verbatim, the sentences about the methods used in the scientific work.

Please don't answer with other text. Just provide the extracted sentences.

**Paragraph:**
"{abstract of the scientific work}"
\end{researchpromptbox}

In the second stage: 
\begin{researchpromptbox}
The following sentences describe the research methods used in a scientific work.
Is there any mention of artificial intelligence related methods in any of these sentences?
Please provide a list of the scientific methods used in the scientific work.
Provide the output in JSON format.
\\
**Sentences:**
"{paragraph}"

**Required JSON Structure:**
- **Answer**: Answer only with Yes or No.
- **Methods**: A comma-separated list of the scientific methods found in the abstract.

**Example of the JSON Output:**

{{
"Answer": "Yes",
"Methods": "{Yes examples}",
}}

**Example of the JSON Output:**
{{
"Answer": "No",
"Methods": "{No examples}",
}}

Please strictly follow the JSON format shown in the examples and
do not add any extra text outside of the JSON structure.
\end{researchpromptbox}
\clearpage
\subsection*{4 - Additional figures:}
\renewcommand{\thefigure}{\arabic{figure}S}
\setcounter{figure}{0}
\begin{figure}[ht]
  \centering
  \includegraphics[width=0.99\linewidth]{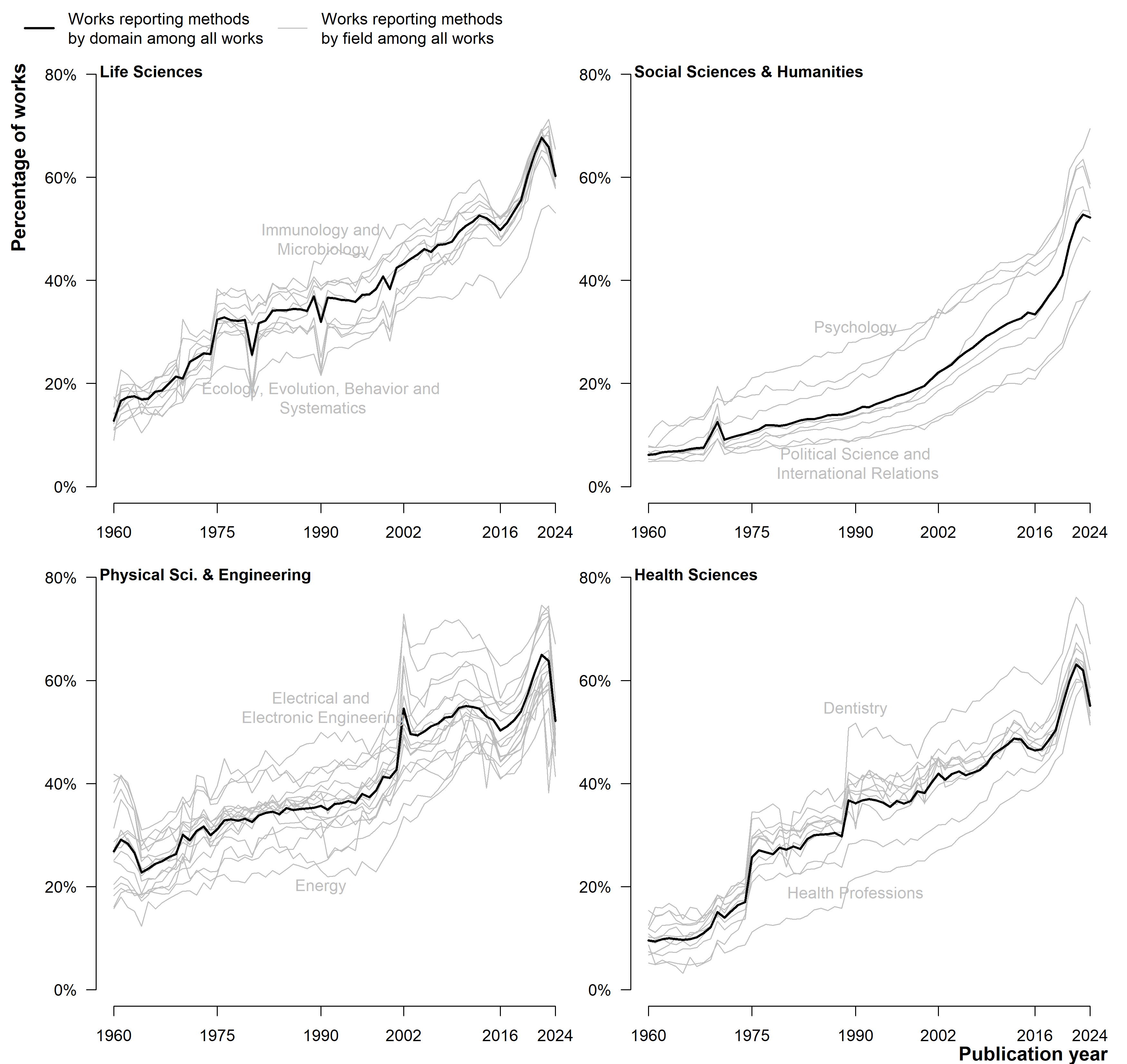}
  \caption{Yearly percentage of academic works reporting any kind of method by scientific domain (black lines) and field (grey lines) as a fraction of all works. Interpretation. Throughout the period of analysis, the number of works reporting any method has grown significantly in all scientific domains, reaching levels of around 50\% in 2024 for all scientific domains and most fields. This figure is high compared to levels as low as, for example, 5\% in 1960 in the Social Sciences and Humanities. This growing trend has to do with greater prevalence of empirical research among all scientific domains and better data quality and standardization of abstract information.}
  \label{fig:feature}
\end{figure}

\begin{figure}[ht]
  \centering
  \includegraphics[width=0.99\linewidth]{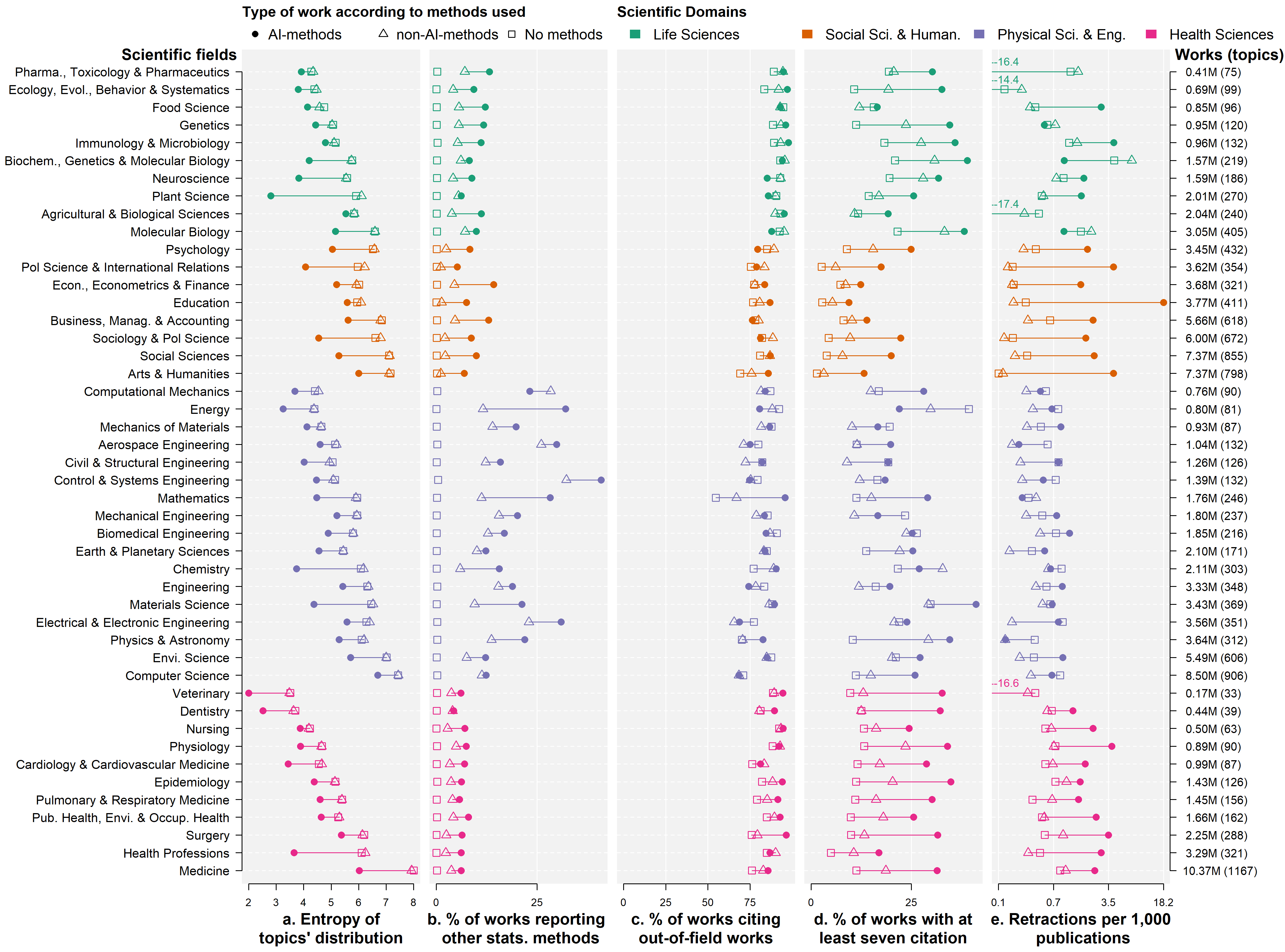}
  \caption{Pre-and post-publication patterns for academic works according to the type of methods: Artificial Intelligence methods (AI-methods), other methods (Methods), and no methods (No methods). The use of methods is determined by the baseline prompting strategy without examples (refer to Materials and Methods sections). Panel a. reports raw mean entropy values. Panels b. to e. report predicted percentages and rates from a multivariate quasi-Poisson model that control for: publication type ("article", "book" , "dissertation", and "preprint"), publication year ("[2000,2005)", "[2005,2010)", "[2010,2015)", "[2015,2020)", "[2020,2025]"), and language ("English", "German",  "Spanish", "French", "Indonesian", "Portuguese", "Other"). Predictions correspond to categories underlined in italic. The analytical sample for all panels except panel (c.) includes 143,232,964 works published between 2002 and 2022. The analytical sample for panel (c.) excludes articles with missing information on cited works (25,994).}
  \label{fig:feature}
\end{figure}

\begin{figure}
  \centering
  \includegraphics[width=0.99\linewidth]{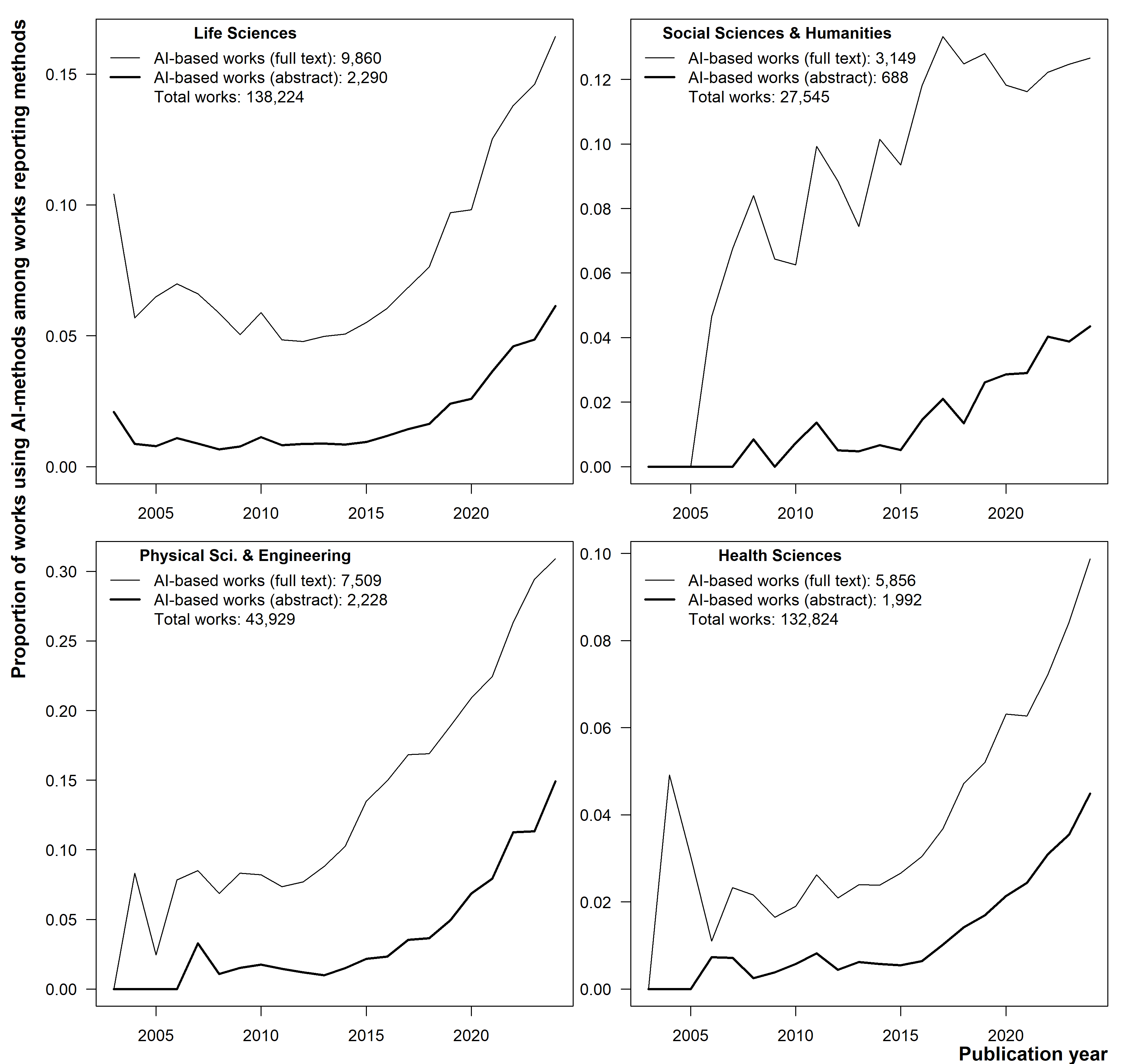}
  \caption{Yearly proportion of academic works using AI methods according to Open Alex abstracts (bold lines) and full texts from PLOS ONE publications from 2003 to 2024 by scientific domains. Source: Open Alex collection for abstracts and domain classification, and PLOS ONE data for full texts (n = 347,522 works)..}
  \label{fig:feature}
\end{figure}

\end{document}